\documentclass[]{IEEEtran}
\usepackage{graphicx,subfigure,epsfig,amsmath,amsbsy,amssymb,citesort,verbatim,url,algorithm,algpseudocode,soul}
\usepackage[usenames]{color}
\usepackage{setspace,amsfonts,epsf,url,bigints}

\def \x {\mathbf{x}}
\def \y {\mathbf{y}}
\def \F {\mathbf{F}}
\def \s {\mathbf{s}}
\def \v {\mathbf{v}}
\def \z {\mathbf{z}}
\def \g {\mathbf{g}}
\def \h {\mathbf{h}}
\def \l {\left}
\def \r {\right}
\def \T {\mathbf{T}}

\def \G {\mathbf{G}}

\def \w {\mathbf{w}}
\def \a {\mathbf{a}}
\def \q {\mathbf{q}}
\def \R {\mathbf{R}}
\newcommand{\sij}[2] {\sum_{#1}^{#2}}
\newcommand{\pij}[2] {\prod_{#1}^{#2}}

\newtheorem{myDef}{Definition}
\newtheorem{myTheorem}{Theorem}
\newtheorem{myLemma}{Lemma}
\newtheorem{myCoro}{Corollary}

\usepackage[normalem]{ulem}

\begin{document}

\title{Performance Limits\\for Noisy Multi-Measurement Vector Problems}
\author{Junan Zhu,~\IEEEmembership{Student Member,~IEEE},
Dror~Baron,~\IEEEmembership{Senior Member,~IEEE}, and Florent Krzakala
\thanks{The work was supported in part by the
National Science Foundation under the Grant CCF-1217749, the
U.S. Army Research Office under the Contract W911NF-14-1-0314, and the European Research
Council under the European Union's 7th Framework Programme (FP/2007-
2013)/ERC Grant Agreement 307087-SPARCS.}
\thanks{Junan Zhu and Dror Baron are with the Department of Electrical and Computer Engineering, NC State University, Raleigh, NC 27695.
E-mail: \{jzhu9, barondror\}@ncsu.edu.}
\thanks{Florent Krzakala is with Sorbonne Universit{\'e}s, Universit{\'e} Pierre et Marie Curie Paris 6 and Ecole Normale Superieure, 75005 Paris, France.
E-mail: florent.krzakala@ens.fr.}}
\maketitle

\begin{abstract}
Compressed sensing (CS) demonstrates that sparse signals can be estimated from under-determined
linear systems. Distributed CS (DCS) further reduces the number of measurements by considering joint sparsity within signal ensembles.
DCS with jointly sparse signals has applications in multi-sensor acoustic sensing, magnetic resonance
imaging with multiple coils, remote sensing, and array signal processing.
Multi-measurement vector (MMV) problems consider the estimation of jointly sparse signals under the  DCS framework. Two related MMV settings are studied. In the first setting, each signal vector is measured by a different independent and identically distributed (i.i.d.) measurement matrix, while in the second setting, all signal vectors are measured by the same i.i.d. matrix.
Replica analysis is performed for these two MMV settings, and the minimum mean squared error (MMSE), which turns out to be identical for both settings, is obtained as a function of the noise variance and number of measurements. To showcase the application of MMV models, the MMSE's of complex CS problems with both real and complex measurement matrices are also analyzed. Multiple performance regions for MMV are identified where the MMSE behaves differently as a function of the noise variance and the number of measurements.

Belief propagation (BP) is a CS signal estimation framework that often achieves the MMSE asymptotically. A phase transition for BP is identified. This phase transition, verified by numerical results, separates the regions where BP achieves the MMSE and where it is suboptimal. Numerical results also illustrate that more signal vectors in the jointly sparse signal ensemble lead to a better phase transition.

\end{abstract}


\section{Introduction}

Compressed sensing (CS)~\cite{CandesRUP,DonohoCS,BaraniukCS2007}
demonstrates that sparse signals can be estimated from under-determined
linear measurements.
Owing to the potential for radically reduced measurement rates, CS has
become an active research area within signal processing.
CS has many application areas including
magnetic resonance imaging~\cite{JuYeKi07,JuSuNaKiYe09},
communication~\cite{Cotter2002scemp}, and
remote sensing~\cite{Ma2009deblur}.

Distributed CS (DCS)~\cite{HN05,DuarteWakinBaronSarvothamBaraniuk2013} is based on the
premise that joint sparsity within signal ensembles enables a further
reduction in the number of measurements.
Motivated by sensor networks~\cite{Pottie2000}, preliminary work in
DCS~\cite{Duarte2006IPSN,HN05,DuarteWakinBaronSarvothamBaraniuk2013} showed that the
number of measurements required per sensor must account for the minimum
features unique to that sensor while features that are common to
multiple sensors are amortized. DCS led to a proliferation
of research on the multi-measurement vector (MMV) problem~\cite{chen2006trs,cotter2005ssl,Mishali08rembo,Berg09jrmm,LeeKimBreslerYe2011,LeeBreslerJunge2012,YeKimBresler2015}.
The MMV problem considers the estimation of a set of sparse signal vectors
that share common supports, and has applications such as radar array signal processing,
acoustic sensing with multiple speakers, magnetic resonance imaging
with multiple coils~\cite{JuYeKi07,JuSuNaKiYe09}, and diffuse optical tomography using multiple
illumination patterns. In MMV, thanks to the common
support, the number of sparse coefficients that can be successfully estimated
increases with the number of
measurements. This property was evaluated rigorously for noiseless
measurements using
$l_0$ minimization~\cite{DuarteWakinBaronSarvothamBaraniuk2013}.
To address measurement noise, estimation approaches for MMV problems
have included greedy algorithms such as SOMP~\cite{tropp2006ass,chen2006trs},
$l_1$ convex relaxation~\cite{malioutov2005ssr,tropp2006ass2}, and M-FOCUSS~\cite{cotter2005ssl}. REduce MMV and BOost (ReMBo) has
been shown to outperform conventional methods~\cite{Mishali08rembo}, and subspace methods have also
been used to solve MMV problems~\cite{LeeBreslerJunge2012,YeKimBresler2015}.
Statistical approaches~\cite{ZinielSchniter2011} often achieve the oracle minimum mean squared error (MMSE).
However, the performance limits of MMV signal estimation in the presence
of measurement noise have not been studied.

Replica analysis is a statistical physics method that can be used to analyze the MMSE and phase transition for inverse problems~\cite{Tanaka2002,GuoVerdu2005,Montanari2006,Krzakala2012probabilistic,krzakala2012statistical,MezardMontanariBook,Barbier2015,Lesieur2015}.
Barbier and Krzakala~\cite{Barbier2015} studied the MMSE for estimating superposition codes using replica analysis. In this paper, we extend the derivation in Barbier and Krzakala~\cite{Barbier2015} to two related yet different MMV settings: ({\em i}) $J$ jointly sparse signals are measured by $J$ different dense matrices that are independent and identically distributed (i.i.d.), and ({\em ii}) $J$ jointly sparse signals are measured by $J$ identical i.i.d. matrices. We only consider dense i.i.d. Gaussian matrices in this work, while our analysis can be extended to other i.i.d. matrices easily.

We make several contributions in this paper. First, we obtain the information theoretic MMSE for the two MMV settings above under the Bayesian setting. Second, we show that in the large system limit the MMSE's for these two settings are identical to the single measurement vector (SMV) problem with a dense measurement matrix and a block sparse signal with fixed length blocks. Third, we derive the MMSE for SMV complex CS problems by noticing that SMV complex CS is essentially an MMV problem. Fourth,
we identify several performance regions for MMV, where the MMSE
has different characteristics based on channel noise variance and measurement rate. Finally, we find a phase transition for belief propagation algorithms (BP)~\cite{DMM2009,CSBP2010,Bayati2011,Montanari2012,Krzakala2012probabilistic,krzakala2012statistical,Barbier2015} applied to MMV problems, which separates regions where BP achieves the MMSE asymptotically and where it is suboptimal. BP simulation results
confirm the phase transition results.

The remainder of the paper is organized as follows.
We introduce our signal and measurement models in Section~\ref{sec:model},
followed by replica analyses for two MMV settings as well as two SMV complex CS problems in Section~\ref{sec:main}. Section~\ref{sec:proof} proves the results of Section~\ref{sec:main}.
Numerical results are discussed in Section~\ref{sec:numeric} and we conclude in Section~\ref{sec:conclusion}.

{\bf Notations:} In this paper, bold capital letters represent matrices, bold lower case letters represent vectors, and normal font lower case letters represent scalars. The entry (scalar) in the $\mu$-th row, $l$-th column of a matrix $\F$ is denoted by $F_{\mu,l}$, where the comma is often omitted. The $\mu$-th entry (scalar) in a vector $\z$ is denoted by $z_{\mu}$.

\section{Signal and Measurement Models}
\label{sec:model}

\setcounter{equation}{0} \indent

{\bf Signal model}:
We consider an ensemble of $J$ signal vectors, $\underline{\s}^j\in\mathbb{R}^N,\ j\in\{1,...,J\}$, where $j$ is the index of the signal.
Consider a {\em super symbol} $\s_l=[\underline{s}_l^1,...,\underline{s}_l^J]^T,\ l\in\{1,...,N\}$, where $[\cdot]^T$ denotes the transpose. The super symbol $\s_l$ follows a $J$-dimensional Bernoulli-Gaussian distribution,
\begin{equation}\label{eq:jsm}
P(\s_l)=\rho \phi(\s_l)+(1-\rho)\delta(\s_l),
\end{equation}
where $\rho$ is the sparsity rate, $\phi(\s_l)$ is a $J$-dimensional Gaussian distribution with zero mean and identity covariance matrix, and $\delta(\s_l)$ is the delta function for $J$-dimensional vectors.

\begin{myDef}[Jointly sparse]\label{def:jointly_sparse}
{\em Ensembles of signals that obey~\eqref{eq:jsm} are called jointly sparse.}
\end{myDef}
{\bf Measurement models}:
Each signal $\underline{\s}^j$ is measured by
an i.i.d. Gaussian measurement matrix $\underline{\F}^j\in\mathbb{R}^{M\times N}$, $\underline{F}_{\mu l}^j \sim \mathcal{N}(0,1/N)$, where $\mu$ refers to the row index and $l$ is the column index. The measurements $\underline{\y}^j$ are corrupted by i.i.d. Gaussian noise $\underline{\z}^j$ consisting of entries $\underline{z}_{\mu}^j\sim \mathcal{N}(0,\Delta)$,
\begin{equation}\label{eq:MMVmodel}
\underline{\y}^j=\underline{\F}^j\underline{\s}^j+\underline{\z}^j,\quad j\in\{1,\cdots,J\}.
\end{equation}
When the number of signal vectors $J=1$, this MMV model~\eqref{eq:MMVmodel} becomes an SMV problem.
Our analyses in this paper are readily extended to other i.i.d. matrices, jointly sparse signals~\eqref{eq:jsm}, and other i.i.d. noise distributions.

\begin{myDef}[MMV-1]\label{def:MMV_set1}
{\em The setting MMV-1 refers to the measurement model in~\eqref{eq:MMVmodel} with all matrices $\underline{F}^j$ being different.}
\end{myDef}
\begin{myDef}[MMV-2]\label{def:MMV_set2}
{\em The setting MMV-2 refers to the measurement model in~\eqref{eq:MMVmodel} with all matrices $\underline{F}^j$ being equal.}
\end{myDef}

In the signal model~\eqref{eq:jsm} and measurement model~\eqref{eq:MMVmodel}, the sparsity rate $\rho$, channel noise variance $\Delta$, and number of channels $J$ are constant.

\begin{myDef}[Large system limit~\cite{GuoWang2008}]\label{def:largeSystemLimit}
The signal length $N$ scales to infinity, and the
number of measurements $M=M(N)$ depends on $N$ and also scales to infinity, where
the ratio approaches a positive constant $R$ for practical problems,
\begin{equation}\label{eq:measurementRate}
\lim_{N\rightarrow\infty} \frac{M(N)}{N} = R>0.
\end{equation}
\end{myDef}
We call $R$ the measurement rate.

\section{Replica Analyses for MMV Settings}
\label{sec:main}
Section~\ref{sec:model} discussed two MMV settings. Both settings have applications in real-world problems such as magnetic resonance imaging~\cite{JuYeKi07,JuSuNaKiYe09} and sensor networks~\cite{Pottie2000}. Although numerous algorithms for MMV signal estimation have been proposed~\cite{tropp2006ass,chen2006trs,malioutov2005ssr,tropp2006ass2,cotter2005ssl,Mishali08rembo,ZinielSchniter2011}, what is missing is an information theoretic analysis of the best possible mean squared error (MSE) performance. Throughout this paper, we only consider the MSE as our performance metric.

\subsection{Statistical physics background and replica method}\label{sec:set1}
\begin{figure}[t]
\centering
\includegraphics[width=8.5cm]{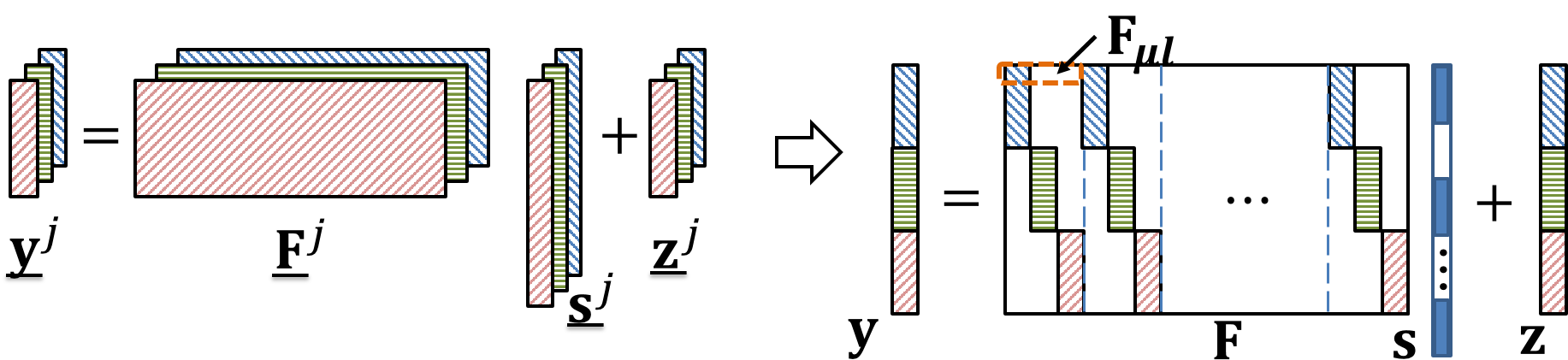}
\caption{Illustration of MMV channel~\eqref{eq:MMVmodel} with $J=3$ signal vectors (left), and one of its possible SMV forms (right). Different background patterns differentiate entries from different channels, and blank space denotes zeros.}\label{fig:channel}
\end{figure}

In order to express~\eqref{eq:MMVmodel} using a single channel, we transform it to an SMV form. One possible way to do so is illustrated in Fig.~\ref{fig:channel}.
The equivalent SMV problem is
\begin{equation}\label{eq:MMVchannel}
  \y=\F\s+\z,
\end{equation}
where $\F\in\mathbb{R}^{MJ\times NJ}$ is the matrix, $\y\in\mathbb{R}^{MJ}$ are the measurements, and the noise is $\z\in\mathbb{R}^{MJ}$.
Entries of the signal vectors $\underline{\s}^j$, measurement vectors $\underline{\y}^j$, and noise vectors $\underline{\z}^j$~\eqref{eq:MMVmodel} form the SMV signal $\s$, measurements $\y$, and noise $\z$~\eqref{eq:MMVchannel} with
\begin{equation*}
s_{(l-1)J+j}=\underline{s}^{j}_l,\ y_{(j-1)M+\mu}=\underline{y}^j_{\mu},\ \text{and}\ z_{(j-1)M+\mu}=\underline{z}^j_{\mu},
\end{equation*}
respectively.
Entries of the matrix $\underline{\F}^j$~\eqref{eq:MMVmodel} form the SMV matrix $\F$~\eqref{eq:MMVchannel} with
$F_{(j-1)M+\mu,(l-1)J+j}=\underline{F}^j_{\mu l}$; other entries of $\F$ are zeros.
The posterior for the estimate $\x\in\mathbb{R}^{NJ}$, comprised of super symbols $\x_l=[x_{(l-1)J+1},...,x_{lJ}]^T,\ l\in\{1,...,N\}$, is
\begin{equation}\label{eq:pxy2}
  P(\x|\y)=\frac{1}{Z}\pij{l=1}{N}P(\x_l)\pij{\mu=1}{MJ}\l[\frac{\operatorname{e}^{-\frac{1}{2\Delta}(\\y_{\mu}-\sij{l=1}{N}\F_{\mu l}\x_l)^2}}{\sqrt{2\pi\Delta}}\r],
\end{equation}
where $\F_{\mu l}=[F_{\mu,(l-1)J+1}, \ldots, F_{\mu,lJ}]$ is a super symbol highlighted by the dashed area in Fig.~\ref{fig:channel}, and the denominator $Z$ is the partition function~\cite{Tanaka2002,GuoVerdu2005,Krzakala2012probabilistic,krzakala2012statistical,MezardMontanariBook,Barbier2015},
\begin{equation}\label{eq:partition}
  Z=\int \pij{l=1}{N}P(\x_l)\pij{\mu=1}{MJ}\l[\frac{\operatorname{e}^{-\frac{1}{2\Delta}(y_{\mu}-\sij{l=1}{N}\F_{\mu l}\x_l)^2}}{\sqrt{2\pi\Delta}}\r]\pij{l=1}{N}d\x_l.
\end{equation}
Note that multi-dimensional integrations such as~\eqref{eq:partition} are denoted by a single $\int$ operator for brevity.
Confining our attention to the Bayesian setting~\cite{Krzakala2012probabilistic,krzakala2012statistical,Barbier2015}, $P(\x_l)$ follows the true distribution~\eqref{eq:jsm},
$P(\x_l)=\rho \phi(\x_l)+(1-\rho)\delta(\x_l)$.

By creating an analogy between the channel~\eqref{eq:MMVchannel} and a many-body thermodynamic system~\cite{Tanaka2002,GuoVerdu2005,Krzakala2012probabilistic,krzakala2012statistical,MezardMontanariBook,Barbier2015}, the posterior~\eqref{eq:pxy2} can be interpreted as the Boltzmann measure on a disordered system with the following Hamiltonian,
\begin{equation}\label{eq:Hamiltonian}
H(\x)=\sum_{l=1}^N \log [P(\x_l)]+\sum_{\mu=1}^{MJ} \frac{1}{2\Delta}\l(y_{\mu}-\sum_{l=1}^N\F_{\mu l} \x_l\r)^2.
\end{equation}

The averaged free energy of the disordered system given by~\eqref{eq:Hamiltonian} characterizes the thermodynamic properties of the system. Evaluating the fixed points (local maxima) in the free energy expression provides the MMSE for the channel~\eqref{eq:MMVchannel}~\cite{Tanaka2002,GuoVerdu2005,Krzakala2012probabilistic,krzakala2012statistical,MezardMontanariBook,Barbier2015}. {\em Under the assumption of self-averaging}~\cite{Tanaka2002,GuoVerdu2005,Krzakala2012probabilistic,krzakala2012statistical,MezardMontanariBook,Barbier2015}, the free energy is defined as\footnote{Part of the literature~\cite{Tanaka2002,GuoVerdu2005} defines the free energy as the negative of~\eqref{eq:free_energy}, so that fixed points of the free energy correspond to local minima.}
\begin{equation}\label{eq:free_energy}
  \mathcal{F}=\lim_{N\rightarrow\infty}\frac{1}{N}\mathbb{E}_{\F,\s,\z}[\log (Z)],
\end{equation}
which is difficult to evaluate. The replica method~\cite{Tanaka2002,GuoVerdu2005,Krzakala2012probabilistic,krzakala2012statistical,MezardMontanariBook,Barbier2015} introduces $n$ replicas of the estimate $\x$ as $\x^a,\ a\in\{1,...,n\}$, and the free energy~\eqref{eq:free_energy} can be approximated by the replica trick~\cite{Krzakala2012probabilistic,krzakala2012statistical,MezardMontanariBook,Barbier2015},
\begin{equation}\label{eq:replicaTrick}
  \mathcal{F}=\lim_{N\rightarrow\infty}\lim_{n\rightarrow 0} \frac{\mathbb{E}_{\F,\s,\z}[ Z^n]-1}{Nn}.
\end{equation}
Note that the self-averaging property that leads to~\eqref{eq:free_energy} and the replica trick~\eqref{eq:replicaTrick}, as well as the replica symmetry assumptions that appear in latter parts of this paper, are assumed to be valid in this work, and their rigorous justification is still an open problem in mathematical physics~\cite{Tanaka2002,GuoVerdu2005,Krzakala2012probabilistic,krzakala2012statistical,MezardMontanariBook,Barbier2015}.\footnote{Recently, the replica Gibbs free
energy has been proven rigorously for the SMV case
by Barbier et al.~\cite{BDMK2016} and
Reeves and Pfister~\cite{ReevesPfister2016}. We conjecture that by generalizing these two works~\cite{BDMK2016,ReevesPfister2016},
our MMV analysis can be made rigorous; we leave it for future work.}

\textbf{Evaluating the free energy}:
To evaluate the free energy~\eqref{eq:replicaTrick}, we calculate $\mathbb{E}_{\F,\s,\z}\l[Z^n\r]$, where $\cdot_{\F,\s,\z}$ denotes expectation with respect to (w.r.t.) $\F,\s$, and $\z$, and $Z$ is given in~\eqref{eq:partition}:
\begin{equation}\label{eq:EZn1}
  \mathbb{E}_{\F,\s,\z}\l[Z^n\r]\!=\!\frac{\mathbb{E}_{\s}\!\l[\displaystyle{\int \pij{l=1}{N}\pij{a=1}{n}\! P(\x_l^a)\!\pij{\mu=1}{M}\!\mathbb{X}_{\mu}\!\pij{l=1}{N}\pij{a=1}{n}d\x_l^a}\r]}{(2\pi\Delta)^{\frac{nMJ}{2}}},
\end{equation}
where
\begin{equation}\label{eq:Xmu}
  \mathbb{X}_{\mu}=\mathbb{E}_{\F,\z}\l[\operatorname{e}^{-\frac{1}{2\Delta}\sij{j=1}{J}\sij{a=1}{n}(v_{\mu j}^a)^2}\r],
\end{equation}
$a$ is the replica index, $\x^a_l$ is the $l$-th super symbol of $\x^a$, and
\begin{equation}\label{eq:v_mu_a}
v_{\mu j}^a=\sij{l=1}{N}\F_{\mu+M(j-1),l}(\s_l-\x_l^a)+z_{\mu+M(j-1)}.
\end{equation}

\begin{myLemma}\label{lemma:covIsSame}
In the large system limit, the quantity $\mathbb{X}_{\mu}$~\eqref{eq:Xmu} is the same for both MMV-1 and MMV-2.
\end{myLemma}

Lemma~\ref{lemma:covIsSame} is proved in Section~\ref{sec:proof}. Because of Lemma~\ref{lemma:covIsSame}, the free energy expressions for MMV-1 and MMV-2 should be identical in the large system limit. We state the result as a theorem and the detailed derivations appear in the Appendix.

\begin{myTheorem}[Free energy for MMV]\label{th:free_energy}
For settings MMV-1 and MMV-2, the free energy expressions as functions of $E$ are identical in the large system limit and are given in~\eqref{eq:free_energy4}.\footnote{The $J$-dimensional integrals in~\eqref{eq:free_energy4} can be simplified to one-dimensional integrals using a change of coordinate to $J$-sphere coordinate. Note also that $E$ approaches the MSE in the large system limit; details appear in the appendix.}
\end{myTheorem}

\begin{figure*}
\begin{eqnarray}
  \mathcal{F}(E)&=&-\frac{J}{2}R\l\{\log[2\pi(\Delta+E)]+\frac{\rho+\Delta}{E+\Delta}\r\}\!+\!\int\!P(\s_1)\! \int\! \log\! \l[ \int P(\x_1)\operatorname{e}^{-\frac{\widehat{Q}+\widehat{q}}{2}\x_1^T\x_1+\widehat{m}\x_1^T\s_1+
  \sqrt{\widehat{q}}\h^T\x_1}\!d\x_1\!\r]\!\mathcal{D}\h \ d\s_1\label{eq:free_energy3}\\
  &=&-\frac{J}{2}R\l\{\log[2\pi(\Delta+E)]+\frac{\Delta}{E+\Delta}\r\}+\frac{JR(1-\rho)}{2(R+E+\Delta)}+\rho\int \log \Bigg[ \rho \l(\frac{E+\Delta}{R+E+\Delta}\r)^{J/2}+\nonumber\\
  &\ &(1-\rho)\operatorname{e}^{-\frac{R}{2(E+\Delta)}\g^T\g}\Bigg]\mathcal{D}\g+(1-\rho)\int \log \l[ \rho \l(\frac{E+\Delta}{R+E+\Delta}\r)^{J/2}+(1-\rho)\operatorname{e}^{-\frac{R}{2(R+E+\Delta)}\h^T\h}\r]\mathcal{D}\h.\label{eq:free_energy4}
\end{eqnarray}
\end{figure*}

\textbf{MMSE}: The $E$ that maximizes the free energy~\eqref{eq:free_energy4} {\em corresponds to} the MMSE~\cite{Krzakala2012probabilistic,krzakala2012statistical,Barbier2015}.
After finding the $E_0$ that maximizes the free energy~\eqref{eq:free_energy4}, we obtain the MMSE, $D_0=E_0$, in the large system limit.

\begin{myCoro}
The MMSE for MMV-1 and MMV-2 is the same for the same measurement rate $R$, noise variance $\Delta$, and number of signal vectors $J$.
\end{myCoro}

{\bf Remark 1:} As the reader can see from the proof of Lemma~\ref{th:free_energy} in Section~\ref{sec:proof}, the key reason that both MMV-1 and MMV-2 have an identical MMSE is that the entries in the super symbols $\s_l$ and $\x_l^{\cdot}$ are i.i.d. That said, we suspect that the MMSE for MMV-1 and MMV-2 could differ by some higher order terms.  If the entries of these super symbols are not i.i.d., which is true in some practical MMV applications~\cite{ZinielSchniter2013MMV}, then it becomes  more difficult to analyze the covariance matrix $\G_{\mu}$ as in Section~\ref{sec:proof}. Therefore, we do not have an analysis for non-i.i.d. entries within $\s_l$ and $\x_l^{\{\cdot\}}$. However, we speculate that MMV-1 might have lower MMSE than MMV-2 in that case.

{\bf Link to SMV with block sparse signal:}
The signal $\s$ in~\eqref{eq:MMVchannel} is a block sparse signal comprised of $N$ blocks of length $J$.
We study a single measurement vector (SMV) problem by replacing the measurement matrix $\F$ in~\eqref{eq:MMVchannel} with an i.i.d. Gaussian matrix ${\mathbf A}\in\mathbb{R}^{MJ\times NJ}$, i.e., $\y={\mathbf A}\s+\z$. The entries of ${\mathbf A}$ follow the distribution, $A_{\mu l}\sim \mathcal{N}(0,\frac{1}{NJ})$. This SMV is similar to the setting in Barbier and Krzakala~\cite{Barbier2015}, except for the different priors and different $\ell_2$ norms in each row of ${\mathbf A}$. We consider these differences while following their derivation~\cite{Barbier2015}, and obtain the same free energy expression as~\eqref{eq:free_energy4}. We have also shown that MMV-1 and MMV-2 have the same MMSE in the large system limit.
Hence, the three settings have the same free energy expression and their MMSE's are the same under the same noise variance $\Delta$ and measurement rate $R$ in the large system limit.

\subsection{Extension to complex SMV}\label{sec:complex}
MMV with jointly sparse signals is a versatile model that can be adapted to other problems. As an example, we show how the MMV model can be used to analyze the MMSE of a complex SMV.\footnote{In Section~\ref{sec:complex}, we only deal with SMV CS, and omit the word ``SMV."}
Consider the complex CS channel, $\y^{\mathcal{C}}=\F^{\mathcal{C}}\s^{\mathcal{C}}+\z^{\mathcal{C}}$,
where $\s^{\mathcal{C}}=\s^{\mathcal{R}}+i\s^{\mathcal{I}}\in\mathbb{C}^N$, $\F^{\mathcal{C}}=\F^{\mathcal{R}}+i\F^{\mathcal{I}}\in\mathbb{C}^{M\times N}$, $\z^{\mathcal{C}}=\z^{\mathcal{R}}+i\z^{\mathcal{I}}\in\mathbb{C}^M$, $\y^{\mathcal{C}}=\y^{\mathcal{R}}+i\y^{\mathcal{I}}\in\mathbb{C}^M$, $i=\sqrt{-1}$, and $\mathcal{R}$ and $\mathcal{I}$ refer to the real and imaginary parts, respectively. The real and imaginary parts of the entries of $\z^{\mathcal{C}}$ both follow a Gaussian distribution, $z_l^{\mathcal{R}}, z_l^{\mathcal{I}}\sim\mathcal{N}(0,\Delta), l\in\{1,...,M\}$.
Assume that the complex signal $\s^{\mathcal{C}}$ is comprised of two jointly sparse signals, $\s^{\mathcal{R}}$ and $\s^{\mathcal{I}}$, that satisfy the $J=2$ dimensional Bernoulli-Gaussian distribution~\eqref{eq:jsm}.
We can extend the analysis of Section~\ref{sec:set1} to two settings of complex CS:
({\em i}) the measurement matrix $\F^{\mathcal{C}}$ is real, and ({\em ii}) $\F^{\mathcal{C}}$ is complex.\footnote{A replica analysis for complex CS with a real measurement matrix appears in Guo and Verd{\'u}~\cite{GuoVerdu2005}. Their derivation does not cover complex matrices.}

{\bf Real measurement matrix:}
Suppose that $\F^{\mathcal{C}}$ is real, $\F^{\mathcal{C}}=\F^{\mathcal{R}}\in \mathbb{R}^{M\times N}$, and the entries of $\F^{\mathcal{R}}$ follow a Gaussian distribution, $F^{\mathcal{R}}_{\mu l}\sim \mathcal{N}(0,\frac{1}{N})$. Complex CS with a real measurement matrix can be written as real-valued MMV,
\begin{equation}\label{eq:complexRealMat}
    \y^{\mathcal{R}}=\F^{\mathcal{R}}\s^{\mathcal{R}}+\z^{\mathcal{R}}\ \text{and}\     \y^{\mathcal{I}}=\F^{\mathcal{R}}\s^{\mathcal{I}}+\z^{\mathcal{I}},
\end{equation}
where $\s^R$ and $\s^I$ are jointly sparse and follow~\eqref{eq:jsm}.
This formulation~\eqref{eq:complexRealMat} fits into MMV-2 for $J=2$. Hence, we can obtain the MMSE according to~\eqref{eq:free_energy4}.\footnote{As a reminder, the free energy of MMV-2 is identical to that of MMV-1 in the large system limit.}

{\bf Complex measurement matrix:}
Consider a complex $\F^{\mathcal{C}}=\F^{\mathcal{R}}+i\F^{\mathcal{I}}\in\mathbb{C}^{M\times N}$ with entries $F_{\mu l}^{\mathcal{R}}, F_{\mu l}^{\mathcal{I}}\sim \mathcal{N}(0,\frac{1}{2N})$.
Expanding out the complex channel, $\y^{\mathcal{C}}=\F^{\mathcal{C}}\s^{\mathcal{C}}+\z^{\mathcal{C}}$, we obtain the equivalent real-valued SMV channel,
\begin{equation}\label{eq.realComplexChannel}
  \begin{bmatrix}
    \y^{\mathcal{R}} \\
    \y^{\mathcal{I}}
  \end{bmatrix}
=
  \begin{bmatrix}
    \F^{\mathcal{R}} & -\F^{\mathcal{I}}  \\
    \F^{\mathcal{I}}  & \F^{\mathcal{R}}
  \end{bmatrix}
    \begin{bmatrix}
    \s^{\mathcal{R}} \\
    \s^{\mathcal{I}}
  \end{bmatrix}
  +
    \begin{bmatrix}
    \z^{\mathcal{R}} \\
    \z^{\mathcal{I}}
  \end{bmatrix}.
\end{equation}

We re-arrange~\eqref{eq.realComplexChannel} as follows,
\begin{equation}\label{eq:complexMatRearrange}
\underbrace{\begin{bmatrix}
    \y^{\mathcal{R}} \\
    \y^{\mathcal{I}}
  \end{bmatrix}}_{\overline{\y}}
\!=\!
\underbrace{\begin{bmatrix}
    \F_{:,1}^{\mathcal{R}},-\F_{:,1}^{\mathcal{I}},...,\F_{:,N}^{\mathcal{R}}, -\F_{:,N}^{\mathcal{I}} \\
    \F_{:,1}^{\mathcal{I}},\ \ \F_{:,1}^{\mathcal{R}},...,\F_{:,N}^{\mathcal{I}},\ \  \F_{:,N}^{\mathcal{R}}
  \end{bmatrix}}_{\overline{\F}}
    \underbrace{\begin{bmatrix}
    s_1^{\mathcal{R}}\\
    s_1^{\mathcal{I}}\\
    \vdots\\
    s_N^{\mathcal{R}}\\
    s_N^{\mathcal{I}}
  \end{bmatrix}}_{\overline{\s}}
  \!+\!
    \underbrace{\begin{bmatrix}
    \z^{\mathcal{R}} \\
    \z^{\mathcal{I}}
  \end{bmatrix}}_{\overline{\z}},
\end{equation}
where $\{:\}$ refers to all the rows.
In the re-arranged channel~\eqref{eq:complexMatRearrange}, the measurement matrix $\overline{\F}$ consists of
super symbols,
\begin{equation}\label{eq:SMV_F}
    \overline{\F}_{\mu l}=\left\{
                \begin{array}{ll}
                 &[F_{\mu l}^{\mathcal{R}},-F_{\mu l}^{\mathcal{I}}],\ \mu\in\{1,...,M\}\\
                 &[F_{\mu l}^{\mathcal{I}}, F_{\mu l}^{\mathcal{R}}],\ \mu\in\{M+1,...,2M\}
                \end{array}
      		\right.,\\
\end{equation}      		
and the signal $\overline{\s}$ consists of
$\overline{\s}_{l}=\begin{bmatrix}
    s_l^{\mathcal{R}} \\
    s_l^{\mathcal{I}}
  \end{bmatrix},\ l\in\{1,...,N\}$.
The measurements and noise are $\overline{\y}=\begin{bmatrix}
    \y^{\mathcal{R}} \\
    \y^{\mathcal{I}}
  \end{bmatrix}$ and
$\overline{\z}=\begin{bmatrix}
    \z^{\mathcal{R}} \\
    \z^{\mathcal{I}}
  \end{bmatrix}$, respectively.
Hence, $\overline{y}_{\mu}=\sum_{l=1}^N \overline{\F}_{\mu l}\overline{\s}_l+\overline{z}_{\mu},\ \mu\in\{1,...,2M\}$.

Section~\ref{sec:proof} shows that the free energy and MMSE for SMV complex CS with complex measurement matrices are the same as MMV-1 with $J=2$.
Note that in the free energy expression~\eqref{eq:free_energy4} of MMV-1, the MSE, $D=E$~\eqref{eq:DandE}, is the average MSE of the $J$ entries of $\s_l$. Therefore, in this complex CS setting, $D$ is the average MSE of the real and imaginary parts of the signal entries.

\section{Proof of Lemma~\ref{lemma:covIsSame}}\label{sec:proof}
In this section, we show that the quantity $\mathbb{X_{\mu}}$ is the same for MMV-1 and MMV-2. Moreover, we show that  complex SMV with a complex measurement matrix also yields the same $\mathbb{X_{\mu}}$ with $J=2$.

First, we re-write~\eqref{eq:Xmu} in the vector form
\begin{equation}\label{eq:XMuVector}
\mathbb{X}_{\mu}\!=\!\mathbb{E}_{\v_{\mu}}\!\left[\operatorname{e}^{-\frac{1}{2\Delta}\sij{j=1}{J}\sij{a=1}{n}(v_{\mu j}^a)^2}\right]\!
=\!\mathbb{E}_{\v_{\mu}}\!\left[\operatorname{e}^{-\frac{1}{2\Delta}\v_{\mu}^T\v_{\mu}}\right],
\end{equation}
where $\v_{\mu}=[v_{\mu 1}^1,...,v_{\mu 1}^a,...,v_{\mu J}^1$, $...,v_{\mu J}^n]^T$ and $v_{\mu j}^a$ is given in~\eqref{eq:v_mu_a}.
In order to calculate the expectation w.r.t. $\v_{\mu}$ in~\eqref{eq:XMuVector}, we calculate the distribution of $\v_{\mu}$, which is approximated by a Gaussian distribution, due to the central limit theorem. The mean is $\mathbb{E}_{\F,\z}[v_{\mu j}^a]=0$.

\begin{figure}[t]
\centering
\includegraphics[width=3.5cm]{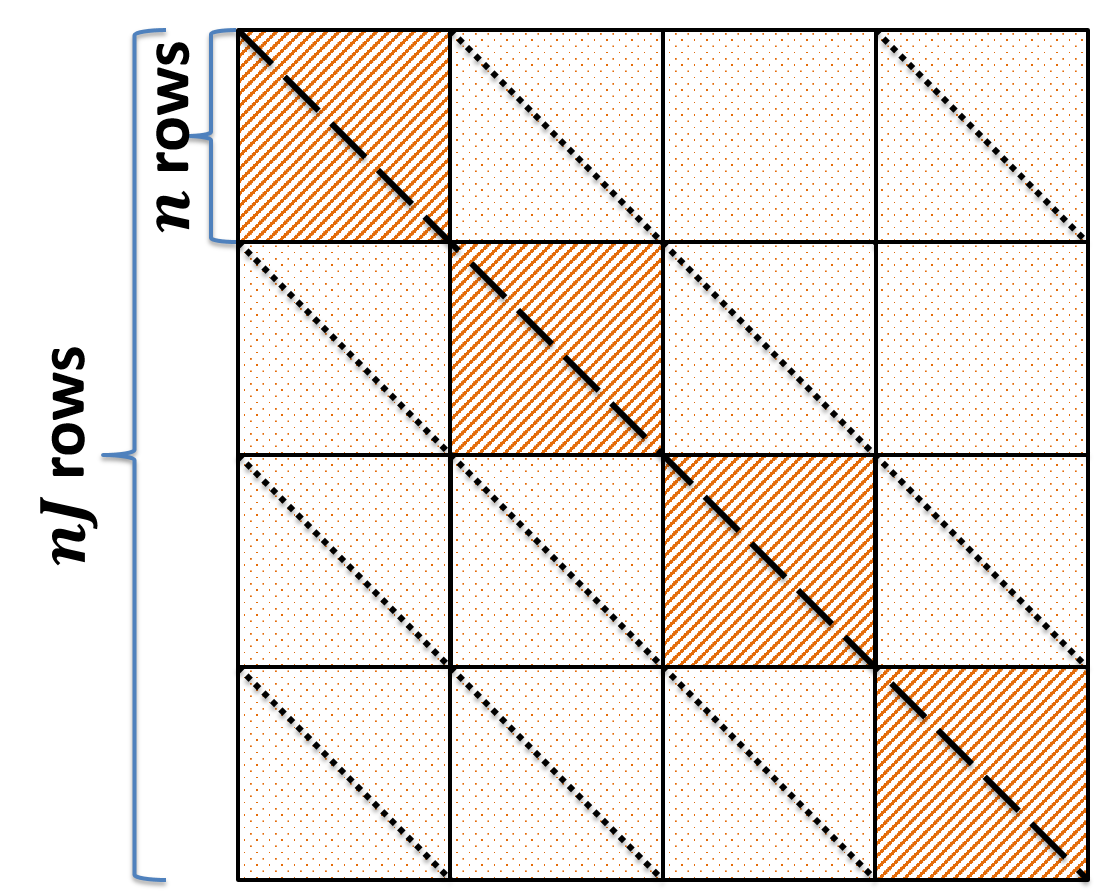}
\caption{Covariance matrix $\G_{\mu}\in\mathbb{R}^{nJ\times nJ}$. Each block in $\G_{\mu}$ has a size of $n\times n$.
The entries in the heavily marked blocks take the value $w_3$, except that entries along the dashed diagonal are $w_1$. The entries in the lightly marked blocks take the value $w_4$, except that entries along the dotted diagonal are $w_2$.}\label{fig.cov}
\end{figure}

We now calculate the covariance matrix, $\G_{\mu}=\mathbb{E}[\v_{\mu}\v_{\mu}^T]$. The matrix is separated into $J\times J$ blocks of size $n\times n$, as shown in Fig.~\ref{fig.cov}. The main diagonal of $\G_{\mu}$ consists of entries $w_1=\mathbb{E}_{\F,\z}[(v_{\mu j}^a)^2]$. The entries in the blocks along the main diagonal (other than entries along the main diagonal itself) are $w_3=\mathbb{E}_{\F,\z}[v_{\mu j}^a v_{\mu j}^b]$. The main diagonals of other blocks have entries $w_2=\mathbb{E}_{\F,\z}[v_{\mu j}^a v_{\mu\eta}^a]$, and other entries in these blocks are $w_4=\mathbb{E}_{\F,\z}[v_{\mu j}^a v_{\mu \eta}^b]$. We now calculate each of these values as follows for MMV-1, MMV-2, and complex SMV with a complex measurement matrix.

{\bf MMV-1:}
We begin by calculating the diagonal entries of the covariance matrix $\G_{\mu}=\mathbb{E}[\v_{\mu}\v_{\mu}^T]$,
\begin{equation}\label{eq:vVar1}
\begin{split}
    &w_1=\mathbb{E}_{\F,\z}\!\l[(v_{\mu j}^a)^2\r]\!=\!\sij{l,k=1}{N,N}\!\Bigg[(\s_l-\x_l^a)^T\times\\
    & \mathbb{E}_{\F}\l\{\F_{\mu+M(j-1), l}^T\F_{\mu+M(j-1), k}\r\}(\s_k-\x_k^a)\Bigg]+\Delta.
\end{split}
\end{equation}
In~\eqref{eq:vVar1}, $\mathbb{E}_{\F}\l\{\F_{\mu+M(j-1), l}^T\F_{\mu+M(j-1), k}\r\}=\frac{\delta_{k,l}}{N}\widetilde{\mathbf{I}}_J$ (cf. Fig.~\ref{fig:channel}), where $\widetilde{\mathbf{I}}_J$ is a $J\times J$ matrix with only one 1 located at row $j$ and column $j$, and $\delta_{k,l}=1$ when $k=l$, else zero. Hence,~\eqref{eq:vVar1} becomes
\begin{eqnarray}
 w_1&=& \mathbb{E}_{\F,\z}\l[(v_{\mu j}^a)^2\r]=\frac{1}{N}\sij{l=1}{N} (s_{l,j}-x_{l,j}^a)^2+\Delta\label{eq:vVar2}\\
  &=&\!\frac{1}{NJ}\!\sij{l=1}{N} \!(\s_l-\x_l^a)^T (\s_l-\x_l^a)\!+\!\Delta,\label{eq:vVar2_1}
\end{eqnarray}
where $s_{l,j}$ and $x_{l,j}^a$~\eqref{eq:vVar2} denote the $j$-th entries in super symbols $\s_l$ and $\x_l^a$, respectively, and~\eqref{eq:vVar2_1} holds because all $J$ entries within the same super symbol ($\s_l$ or $\x_l^a$) are i.i.d.

Similarly, we obtain
\begin{equation}
\begin{split}
w_2=&\mathbb{E}_{\F,\z}[v_{\mu j}^a v_{\mu\eta}^{a}] = \frac{1}{N}\sij{l=1}{N}(s_{l,j}-x_{l,j}^a)(s_{l,\eta}-x_{l,\eta}^{a})\\
     &= \frac{1}{NJ}\sij{l=1}{N}(\s_l-\x_l^a)^T (\s_l^a-\x_l^b),\label{eq:sx_iid1}
\end{split}
\end{equation}
where entries of $\s_l^a$ follow the same distribution as entries of $\s_l$ given $l$, and~\eqref{eq:sx_iid1} is due to
({\em i}) entries of $\s_l$ being i.i.d., ({\em ii}) entries of $\x_l^{\{\cdot\}}$ being i.i.d. for fixed $l$, and ({\em iii}) the replica symmetry assumption~\cite{Krzakala2012probabilistic,krzakala2012statistical}.
We also obtain
\begin{equation}\label{eq:v_j_eta_a}
\begin{split}
w_3=\mathbb{E}_{\F,\z}[v_{\mu j}^a v_{\mu j}^b]\!&=\!\frac{1}{NJ}\!\sij{l=1}{N}(\s_l\!-\!\x_l^a)^T \!(\s_l-\x_l^b)+\Delta.\\
w_4=\mathbb{E}_{\F,\z}[v_{\mu j}^a v_{\mu\eta}^{b}]&=\frac{1}{NJ}\sij{l=1}{N}(\s_l-\x_l^a)^T (\s_l^a-\x_l^b),
\end{split}
\end{equation}

We now define the following auxiliary parameters
\begin{equation}\label{eq:auxParamsSet1}
\begin{split}
m_a=\frac{\displaystyle\sij{l=1}{N} (\x_l^a)^T\s_l}{NJ},&\quad Q_a=\frac{\displaystyle\sij{l=1}{N} (\x_l^a)^T\x_l^a}{NJ}, \\
q_{ab}=\frac{\displaystyle\sij{l=1}{N} (\x_l^a)^T\x_l^b}{NJ},&\quad
q_0=\frac{1}{NJ}\sij{l=1}{N}(\s_l^a)^T \s_l,
\end{split}
\end{equation}
which allow us to express \eqref{eq:vVar2_1}--\eqref{eq:v_j_eta_a} as
\begin{equation*}
w_1=\rho-2m_a+Q_a+\Delta,
\end{equation*}
\begin{equation}\label{eq:ws2}
w_2=q_0-(m_a+m_b)+q_{ab},
\end{equation}
\begin{equation*}
w_3 = \rho-(m_a+m_b)+q_{ab}+\Delta,
\end{equation*}
\begin{equation}\label{eq:ws4}
w_4=q_0-(m_a+m_b)+q_{ab}.
\end{equation}

Plugging the distribution of $\v_{\mu}$, approximated by $P(\v_{\mu})=[(2\pi)^n\det (\G_{\mu})]^{-\frac{1}{2}}\exp(-\frac{1}{2}\v_{\mu}^T\G_{\mu}^{-1}\v_{\mu})$, into~\eqref{eq:XMuVector}, we obtain
\begin{equation}\label{eq:Xmu2}
\mathbb{X}_{\mu}=\l[\det(\mathbb{I}_n+\frac{1}{\Delta}\G_{\mu})\r]^{-1/2}.
\end{equation}

{\bf MMV-2:}
For the matrix $\F$~\eqref{eq:MMVchannel} in this setting, rows $jM+1,...,(j+1)M,\ 2\leq j \leq J$, will be the right-shift of rows $(j-1)M+1,...,jM$.
We express $v_{\mu j}^a$~\eqref{eq:v_mu_a} as
\begin{equation}\label{eq:v_mu_j_a_mmv2}
  v_{\mu j}^a=\sij{l=1}{N}\F_{\mu l}{\bf T}_{j}(\s_l-\x_l^a)+z_{\mu+M(j-1)},\ \mu\in \{1,\cdots,M\},
\end{equation}
where $\T_{j}$ is a $J\times J$ transform matrix with the $j$-th entry of the first row being one and all other entries in $\T_j$ being zeros. Using the same derivations as in MMV-1, it can be proved that the covariance matrix $\G_{\mu}=\mathbb{E}[\v_{\mu}\v_{\mu}^T]$  in MMV-2 is identical to that of MMV-1.
Therefore, $\mathbb{X}_{\mu}$ in MMV-1 and MMV-2 are identical in the large system limit.

{\bf Complex SMV with complex measurement matrix:}
The derivations are the same as in MMV-2 above, except that we need to change $\F_{\mu l}$
in~\eqref{eq:v_mu_j_a_mmv2}  to $\overline{\F}_{\mu l}$~\eqref{eq:SMV_F} and
replace $\T_j$ by
\begin{equation}\label{eq:TransMat}
{\bf T}=\begin{bmatrix}
    0 & 1 \\
    -1 & 0
  \end{bmatrix},
\end{equation}
because $\overline{\F}_{(\mu+M)l}=\overline{\F}_{\mu l}{\bf T},\ \mu\in\{1,...,M\}$.
Using similar steps as above, we obtain that the covariance matrix $\G_{\mu}$ in this case, is also the same as that of MMV-1 with $J=2$.

\textbf{Solving $\mathbb{X}_{\mu}$}: For such a structured matrix $\G_{\mu}$ (Fig.~\ref{fig.cov}), elementary transforms show that the eigen-values (EV's) are comprised of one EV equal to $\alpha_1=[w_1+(J-1)w_2]+(n-1)[w_3+(J-1)w_4],\ (J-1)$ EV's equal to $\alpha_2=(w_1-w_2)+(n-1)(w_3-w_4),\ (n-1)$ EV's equal to $\alpha_3=[w_1+(J-1)w_2]-[w_3+(J-1)w_4]$, and $(J-1)(n-1)$ EV's equal to $\alpha_4=(w_1-w_2)-(w_3-w_4)$.

Owing to replica symmetry~\cite{Krzakala2012probabilistic,krzakala2012statistical}, we have $m_a=m_b=m$, $Q_a=Q$, and $q_{ab}=q$. Also, in the Bayesian setting, we have $m=q_0=q$ and $Q=\rho$.
Thus, $w_2=w_4=0$ (\eqref{eq:ws2} and~\eqref{eq:ws4}),
and
\begin{equation}\label{eq:detSet2}
\begin{split}
\det (\mathbb{I}_{nJ}+&\frac{1}{\Delta}\G_{\mu})= \l(1+\frac{\alpha_1}{\Delta}\r)\l(1+\frac{\alpha_2}{\Delta}\r)^{J-1}\times\\
&\quad \l(1+\frac{\alpha_1}{\Delta}\r)^{n-1}\l(1+\frac{\alpha_1}{\Delta}\r)^{(n-1)(J-1)}\\
&=\l(1+n\frac{w_3}{\Delta+\alpha_4}\r)^J\!\l(1+\frac{1}{\Delta}\alpha_4\r)^{Jn}\!.
\end{split}
\end{equation}
Considering~\eqref{eq:detSet2}, we simplify~\eqref{eq:Xmu2},
\begin{equation}\label{eq:XmuNew}
\lim_{n\rightarrow 0}\mathbb{X}_{\mu}=\operatorname{e}^{-\frac{nJ}{2}\l[\frac{\rho-2m+\Delta+q}{Q-q+\Delta}+\log(Q-q+\Delta)-\log(\Delta)\r]},
\end{equation}
where we rely on the following Taylor series,
\begin{equation}\label{eq:firstOrder1}
\operatorname{e}^{nk}\approx 1+nk\Rightarrow \operatorname{e}^{-\frac{n}{2}k}\approx (1+nk)^{-1/2},\ n\rightarrow 0.
\end{equation}

\section{Numerical Results}\label{sec:numeric}
Given a free energy expression for a CS problem, the MMSE can be obtained by evaluating the largest free energy~\cite{Tanaka2002,GuoVerdu2005,Krzakala2012probabilistic,krzakala2012statistical,MezardMontanariBook,Barbier2015}.  Having derived the free energy for the two  MMV settings in Section~\ref{sec:main}, this section calculates the MMSE under various cases. Different performance regions of MMV are identified, where the MMSE behaves differently as a function of the noise variance $\Delta$ and measurement rate $R$. We identify a phase transition of belief propagation (BP) that separates regions where BP is optimal asymptotically or not. Simulation results match the predicted performance of BP.

\subsection{Performance regions: Definitions and numerical results}\label{sec:PerfRegion}
\begin{figure}[t] 
\centering
\includegraphics[width=8cm]{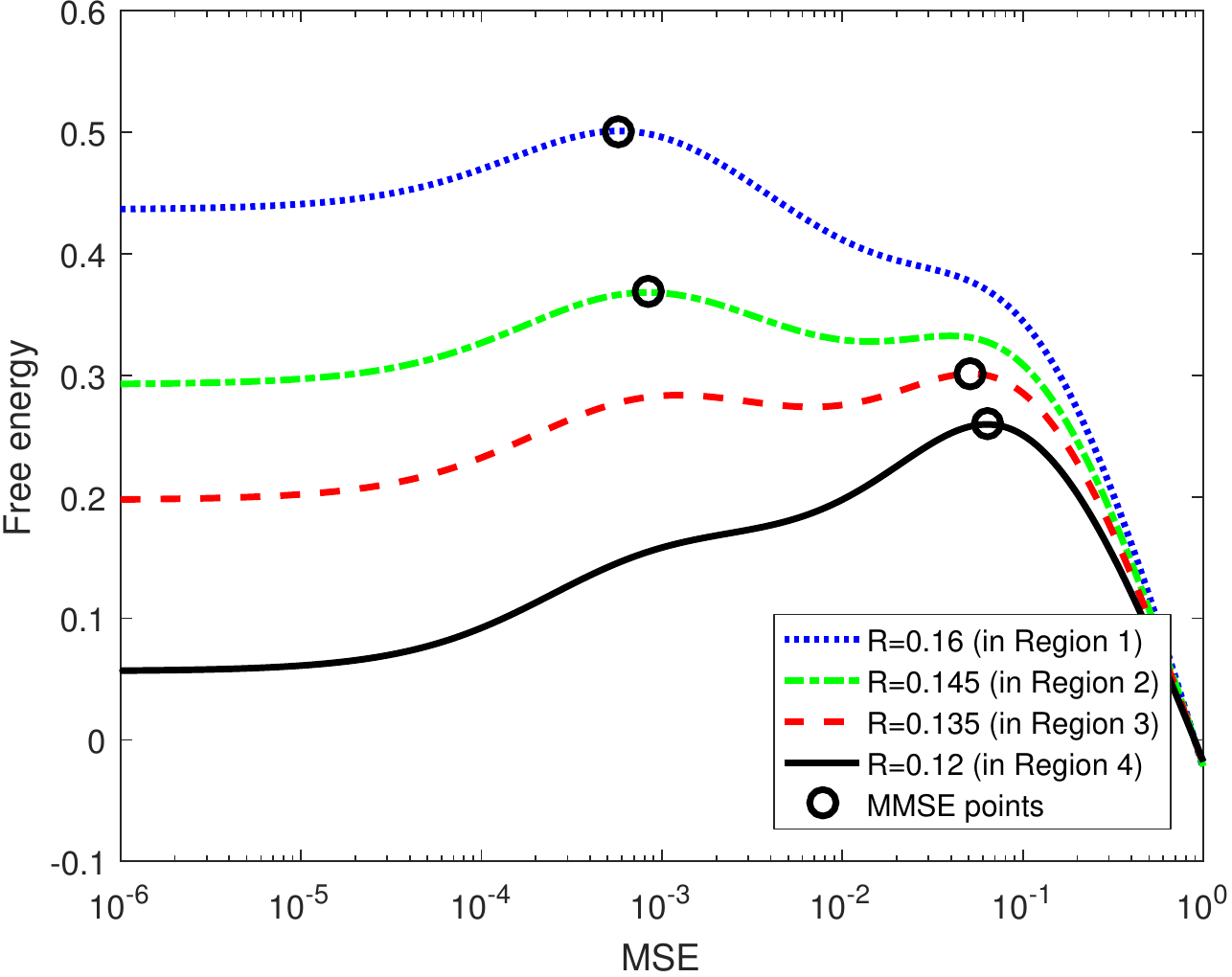}
\caption{Free energy as a function of MSE for different measurement rates $R$  (number of jointly sparse signal vectors $J=3$ and noise variance $\Delta=-35$ dB). The black circles mark the largest free energy, and so they correspond to the MMSE.}\label{fig:freeEnergyProf}
\end{figure}

When calculating the MMSE~\eqref{eq:DandE} for different settings from the free energy expression~\eqref{eq:free_energy4}, four different {\em performance regions} will appear, as  illustrated in Fig.~\ref{fig:freeEnergyProf} and discussed below.

{\bf Regions 1 and 4:} The free energy~\eqref{eq:free_energy4} has one local maximum point w.r.t. the MSE $D$~\eqref{eq:DandE}. This $D$ leads to the globally maximum free energy and is the MMSE.

{\bf Regions~2 and~3:} There are 2 local maxima in the free energy, $D_1$ and $D_2$, where $D_1<D_2$. In Region~2, the smaller MSE, $D_1$, leads to the larger local maximum free energy~\eqref{eq:free_energy4} (hence, $\mathcal{F}(D_1)$ is the global maximum), and is the MMSE. In Region~3, the larger MSE, $D_2$, is the MMSE.

{\bf Boundaries between regions:} We denote the boundary separating regions~1 and~2 by the {\em BP threshold} $R_{BP}(\Delta)$, the boundary separating regions~2 and~3 by the {\em low noise threshold} $R_l(\Delta)$, and  the boundary separating regions~3 and~4 by the {\em critical threshold} $R_c(\Delta)$.

\begin{figure*}[t]
  \subfigure[]{
    \label{fig:MMV_J1} 
    \includegraphics[width=5.8cm]{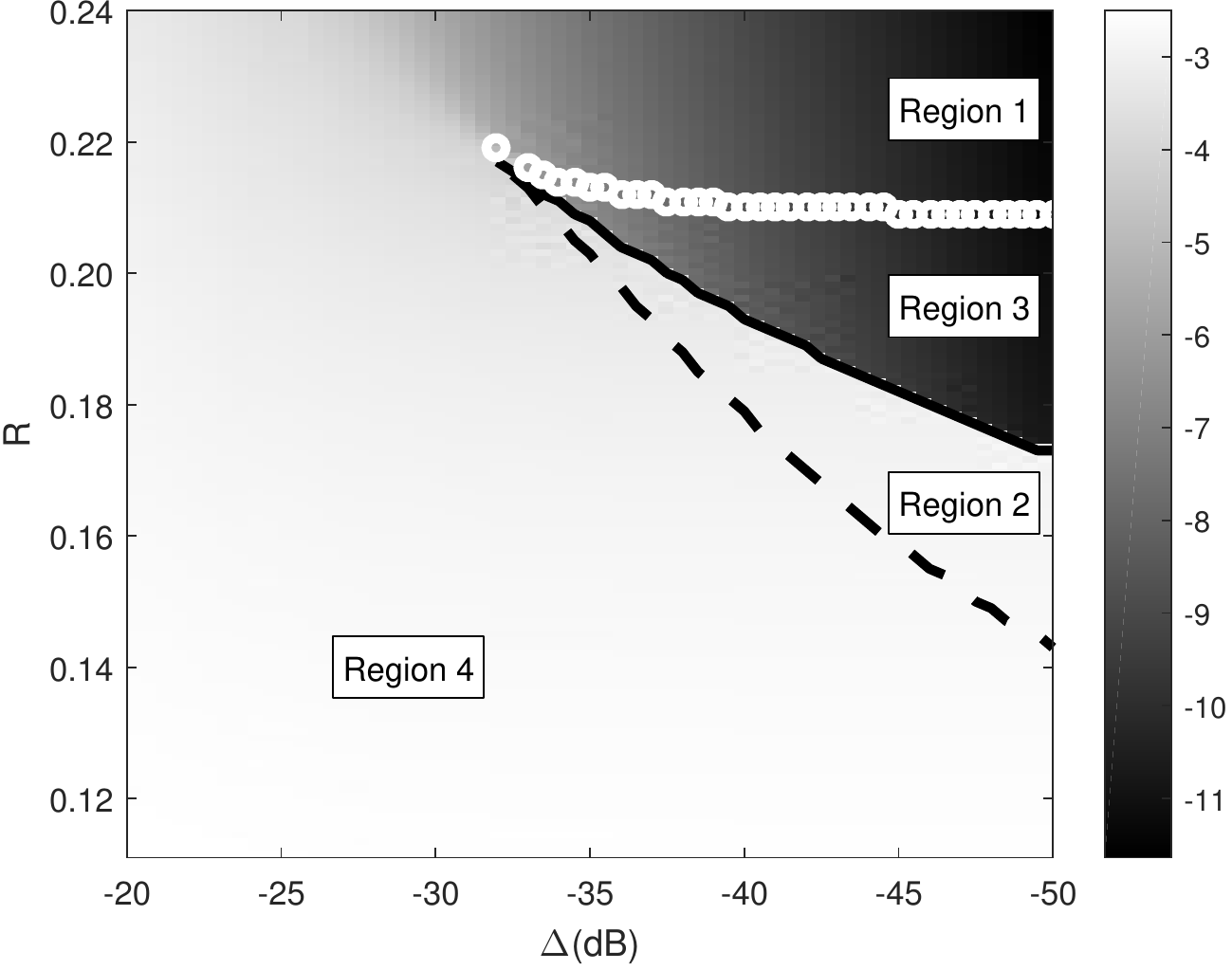}}
  \subfigure[]{
    \label{fig:MMV_J3} 
    \includegraphics[width=5.8cm]{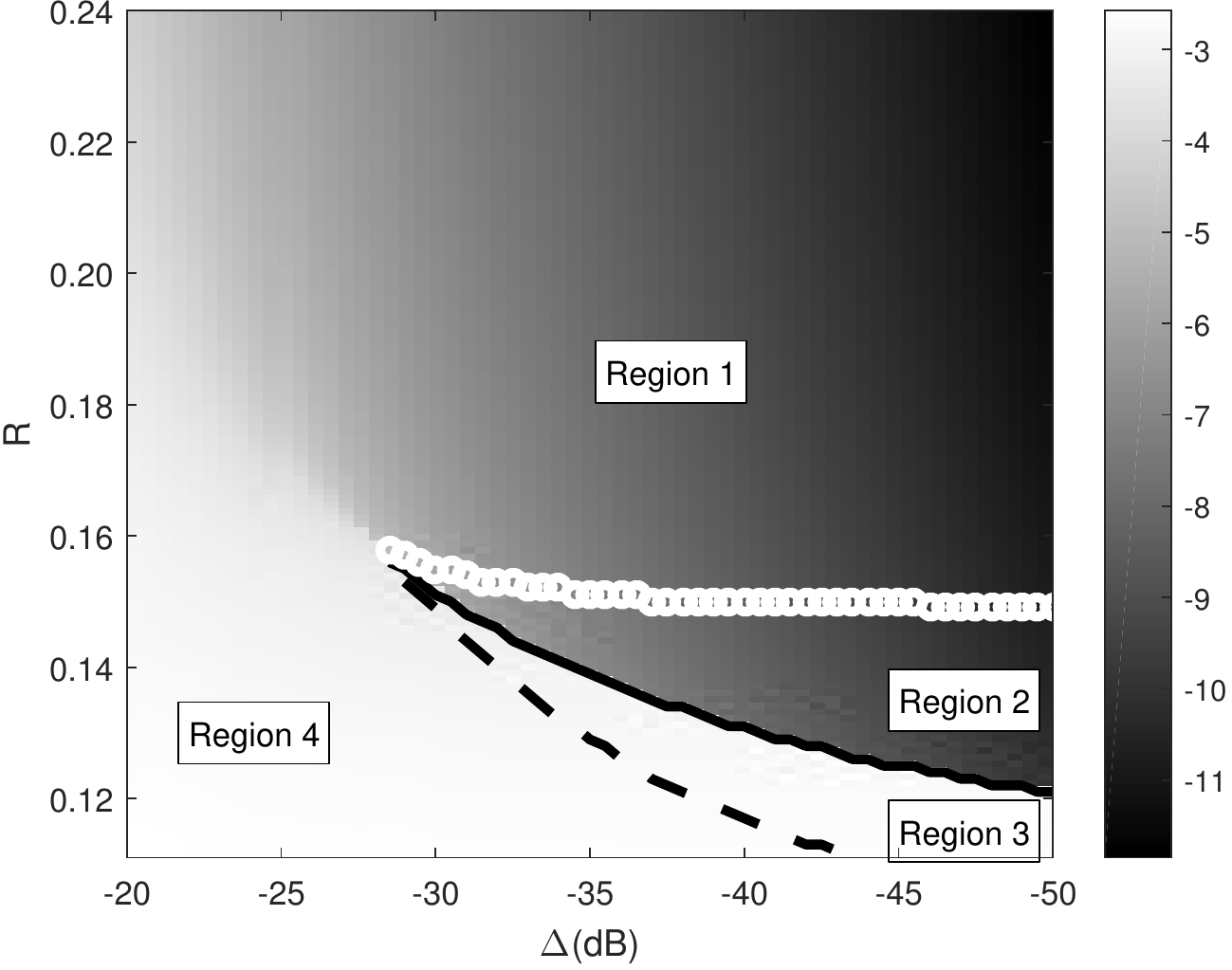}}
  \subfigure[]{
    \label{fig:MMV_J5} 
    \includegraphics[width=5.8cm]{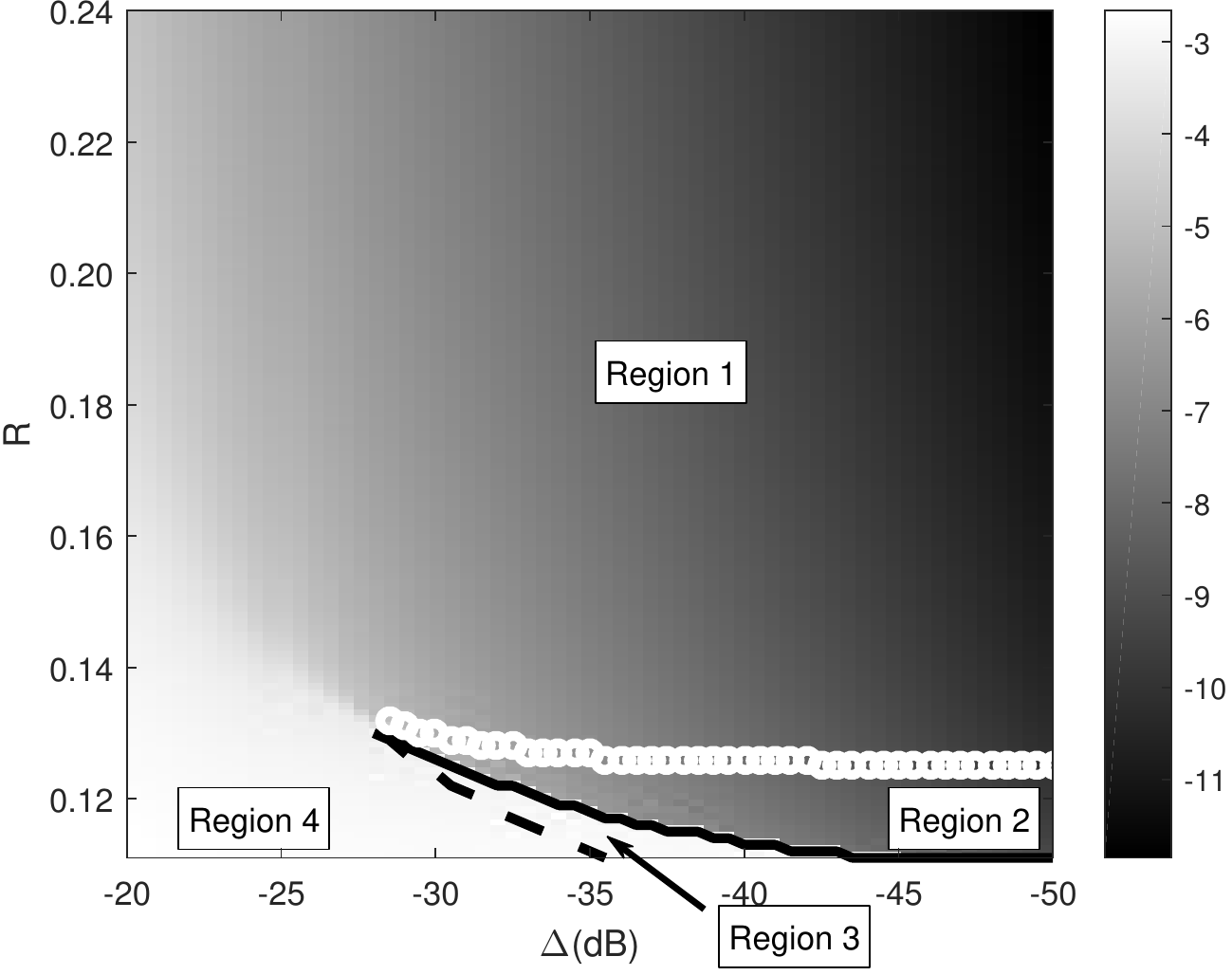}}
\caption{Performance regions for MMV with different $J$. The darkness of the shades corresponds to $\ln$(MMSE) for a certain noise variance $\Delta$ and measurement rate $R$. There are 4 regions, Regions~1 to~4, where the MMSE as a function of the noise variance $\Delta$ and measurement rate $R$ behaves differently. Regions~1 to~4 are separated by 3 thresholds, $R_c(\Delta)$ (the dashed curves), $R_l(\Delta)$ (the solid curves), and $R_{BP}(\Delta)$ (the curves comprised of little white circles); note that Section~\ref{sec:PerfRegion} discusses how to obtain these thresholds. (a) MMV with $J=1$, (b) MMV with $J=3$, and (c) MMV with $J=5$.}\label{fig:PerformanceRegions}
\end{figure*}

{\bf Numerical results:}
Consider $J$-dimensional Bernoulli-Gaussian signals~\eqref{eq:jsm} with sparsity rate $\rho=0.1$.
Evaluating the free energy~\eqref{eq:free_energy4} with the noise variance $\Delta$ from -20 dB to -50 dB and measurement rate $R$ from 0.11 to 0.24, we obtain the MMSE as a function of $\Delta$ and $R$ for $J=1,3$, and $5$, as shown in Fig.~\ref{fig:PerformanceRegions}.\footnote{The MMV with $J=1$ becomes an SMV. The MMSE results in Fig.~\ref{fig:MMV_J1} match with the SMV MMSE in Krzakala et. al.~\cite{Krzakala2012probabilistic,krzakala2012statistical} and Zhu and Baron~\cite{ZhuBaronCISS2013}.} The darkness of the shades represents the natural logarithm of the MMSE, $\ln$(MMSE). In all panels,
the critical threshold $R_c(\Delta)$, low noise threshold $R_l(\Delta)$, and BP threshold $R_{BP}(\Delta)$, as well as Regions~1-4, are marked.

In Regions~3 and~4, the best-possible algorithm yields a large MMSE for all noise variances. In contrast, in Regions~1 and~2, the optimal algorithm yields an MMSE that decreases with the noise variance $\Delta$. To summarize, the optimal algorithm yields poor estimation performance below the low noise threshold $R_{l}(\Delta)$, and good performance above $R_{l}(\Delta)$.

\begin{figure}[t] 
\centering
\includegraphics[width=8cm]{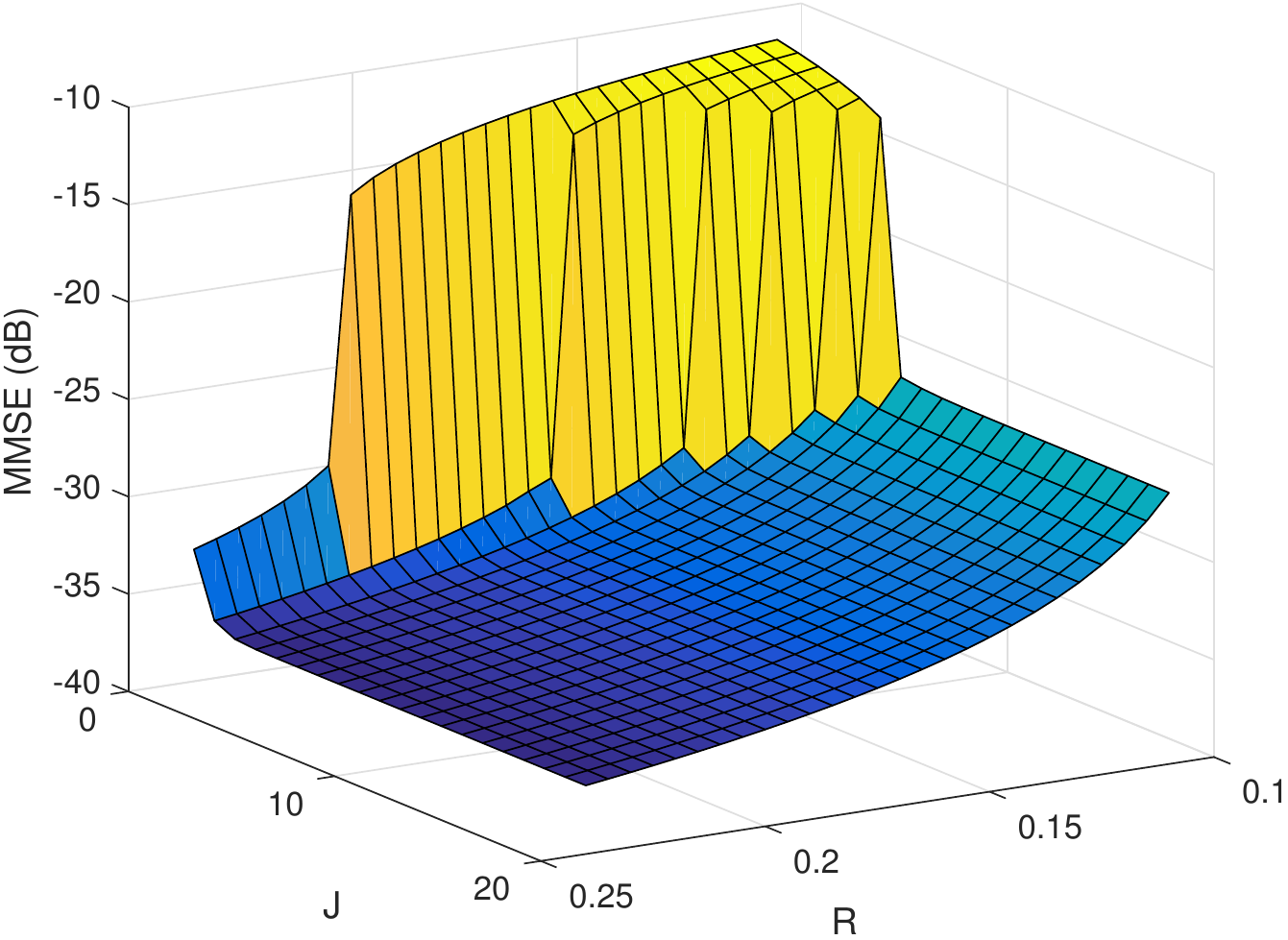}
\caption{MMSE in dB as a function of measurement rate $R$ and number of jointly sparse signal vectors $J$ (noise variance $\Delta=-35$ dB).}\label{fig:MMSE_R_J}
\end{figure}

We further examine the MMSE as a function of the number of jointly sparse signal vectors $J$ and the measurement rate $R$. We plotted the MMSE in dB scale in Fig.~\ref{fig:MMSE_R_J}. The noise variance is -35 dB. We can see that the MMSE decreases with more signal vectors $J$ and greater measurement rate $R$. However, the MMSE depends less on $J$ as $J$ is increased. Note that the discontinuity in the MMSE surface in Fig.~\ref{fig:MMSE_R_J} is a result of the different performance regions that the various settings (different $J$ and $R$) lie in.

\subsection{BP phase transition}\label{sec:phaseTrans}
Belief propagation (BP)~\cite{DMM2009,CSBP2010,Montanari2012,Bayati2011,Krzakala2012probabilistic,krzakala2012statistical,Barbier2015} is an algorithmic framework motivated by statistical physics, which can often achieve the optimal estimation performance (MMSE). When there are multiple local maxima $D_1<D_2$ in the free energy~\eqref{eq:free_energy4}, BP converges to the local maximum with the larger MSE, $D_2$~\cite{DMM2009,Montanari2012,Bayati2011,Krzakala2012probabilistic,krzakala2012statistical}. Hence, $D_2$ characterizes the {\em predicted MSE} for BP. Moving from Region~1 to Region~2 by decreasing the measurement rate $R$ with fixed noise variance $\Delta$, the number of local maxima increases from 1 to 2. Therefore, BP estimation performance experiences a sudden deterioration (increase in MSE) when the measurement rate $R$ drops such that the combination of the noise variance $\Delta$ and measurement rate $R$ moves from Region~1 to Region~2. The BP threshold, $R_{BP}(\Delta)$, is the boundary between Regions~1 and~2, and is where the BP phase transition happens. That is, BP achieves poor estimation performance below $R_{BP}(\Delta)$, and good performance above $R_{BP}(\Delta)$.

{\bf Remark 2}: In Fig.~\ref{fig:PerformanceRegions}, we see that increasing $J$ reduces the BP threshold $R_{BP}(\Delta)$. Since BP achieves the MMSE when $R>R_{BP}(\Delta)$, increasing $J$ is beneficial to applications that use BP as the estimation algorithm.

{\bf Remark 3}: We further analyzed the low noise ($\Delta\rightarrow 0$) and zero noise ($\Delta=0$) cases. The critical threshold $R_c(\Delta)$ converges to $\rho$ as the noise variance $\Delta$ is decreased for $J=1,3$, and $5$. We believe that this numerical result holds for every $J$. Moreover, this result matches the theoretical robust threshold of Wu and Verd{\'u}~\cite{WuVerdu2012} for $J=1$ in the low noise limit. Our numerical results also show that the BP threshold $R_{BP}(\Delta)$ converges to some value for different $J$ as $\Delta\rightarrow 0$. Analyzing these observations rigorously is left for future work.

\subsection{BP simulation}\label{sec:AMPsim}

After obtaining the theoretic MMSE for MMV, as well as the predicted MSE for BP, we run some simulations to estimate the $\underline{\s}^j$ of channel~\eqref{eq:MMVmodel} in a Bayesian setting.
The algorithm we use is approximate message passing (AMP)~\cite{DMM2009,Montanari2012,Bayati2011,Krzakala2012probabilistic,krzakala2012statistical,Barbier2015}, which is an approximation to the BP algorithm; related algorithms have been proposed by Ziniel and Schniter~\cite{ZinielSchniter2013MMV} and Kim et al.~\cite{KimChangJungBaronYe2011}.
In the SMV case, when the measurement matrix and the signal have i.i.d. entries, AMP has the state-evolution (SE) property~\cite{DMM2011,Bayati2011,JavanmardMontanari2012,Donoho2013,Bayati2015} that tracks the evolution of the MSE at each iteration. Recently, Javanmard and Montanari proved that SE tracks AMP rigorously in an SMV setting with a spatially coupled measurement matrix~\cite{JavanmardMontanari2012}. According to our transform in Fig.~\ref{fig:channel}, we can see that the proof~\cite{JavanmardMontanari2012} could be extended to the MMV setting. Note that SE allows to compute the
highest equilibrium of Gibbs free energy~\cite{DMM2011,Bayati2011,JavanmardMontanari2012,Donoho2013,Bayati2015}, which corresponds to the local optimum $D_2$ in Section~\ref{sec:phaseTrans}. Hence, AMP often achieves the same MSE as BP
and we use AMP simulation results to demonstrate that the MMSE can often be achieved.\footnote{When the assumptions about the measurement matrix and signal~\cite{DMM2009,Montanari2012,Bayati2011,Krzakala2012probabilistic,krzakala2012statistical,Barbier2015} are violated, AMP might suffer from convergence issues.}
Considering the structure of $\F$, we simplify the AMP algorithm in Barbier and Krzakala~\cite{Barbier2015} to obtain Algorithm~\ref{algo:AMP_MMV},\footnote{Note that Algorithm~\ref{algo:AMP_MMV} is a straightforward simplification of the AMP algorithm in Barbier and Krzakala~\cite{Barbier2015}.} where $\{\Sigma_j\}_{j=1}^J$, $\{R^j_l\}_{j=1}^J$, $\{a_l^j\}_{j=1}^J$ and $\{v_l^j\}_{j=1}^J$ refer to sets of all intermediate variables $\Sigma_j$, pseudodata $R^j_l$, estimates $a_l^j$, and variances $v^j_l,\ j\in\{1,...,J\},\ l\in\{1,...,N\}$, respectively.
The current iteration $t$, change in the estimate $\delta$, and intermediate variables $\Theta_j,\ j\in\{1,...,J\}$, are scalars. The intermediate variables $\q^j$ and $\w^j$ are vectors of length $M$. The functions $f_{a_l}(\{\Sigma_j\}_{j=1}^J,\{R^j_l\}_{j=1}^J)$ and $f_{v_l}(\{\Sigma_j\}_{j=1}^J,\{R^j_l\}_{j=1}^J)$ are given by
\begin{equation*}
\begin{split}
&f_{a_l}(\{\Sigma_j\}_{j=1}^J,\{R^j_l\}_{j=1}^J)=\\
&\frac{\rho\frac{1}{\Sigma_j+1}\{R^j_l\}_{j=1}^J}{\rho+(1-\rho)\pij{j=1}{J}\l\{\sqrt{1+\frac{1}{\Sigma_j}}\exp\l[-\frac{(R^j_l)^2}{2\Sigma_j(\Sigma_j+1)}\r]\r\}},
\end{split}
\end{equation*}
\begin{equation*}
\begin{split}
&f_{v_l}(\{\Sigma_j\}_{j=1}^J,\{R^j_l\}_{j=1}^J)=-\l[f_{a_l}(\{\Sigma_j\}_{j=1}^J,\{R^j_l\}_{j=1}^J)\r]^2\\
&+\frac{\rho\frac{1}{\Sigma_j+1}\l[(\{R^j_l\}_{j=1}^J)^2\frac{1}{\Sigma_j+1}+\Sigma_j\r]}{\rho+(1-\rho)\pij{j=1}{J}\l\{\sqrt{1+\frac{1}{\Sigma_j}}\exp\l[-\frac{(R^j_l)^2}{2\Sigma_j(\Sigma_j+1)}\r]\r\}},
\end{split}
\end{equation*}
for $J$-dimensional Bernoulli-Gaussian signals~\eqref{eq:jsm}.

\begin{algorithm}[t]
\caption{AMP for MMV}
\label{algo:AMP_MMV}
\begin{algorithmic}[1]
\\{\bf Inputs:} Maximum number of iterations $t_{max}$, threshold $\epsilon$, sparsity rate $\rho$, noise variance $\Delta$, measurements $\y^j$, and measurement matrices $\F^j, \forall j$
\\{\bf Initialize:} $t=1,\delta=\infty,\w^j=\y^j,\Theta_j=0,v^j_l=\rho\Delta,a^j_l=0,\forall l,j$
\While{$t<t_{max}$ and $\delta>\epsilon$}
\For{$j\leftarrow 1$ to $J$}
\\\quad\quad\quad$\q^j=\frac{\y^j-\w^j}{\Delta+\Theta_j}$
\\\quad\quad\quad$\Theta_j=\frac{1}{N}\sum_{l=1}^N v^j_l$
\\\quad\quad\quad$\w^j=\F^j \a^j-\Theta_j \q^j$
\\\quad\quad\quad$\Sigma_j=\frac{N(\Delta+\Theta_j)}{M}$ // Scalar channel noise variance
\\\quad\quad\quad$\R^j=\a^j+\Sigma_j (\F^j)^T \frac{\y^j-\w^j}{\Delta+\Theta_j}$ // Pseudodata
\\\quad\quad\quad$\widehat{\a}^j=\a^j$ // Save current estimate
\EndFor
\For{$l\leftarrow 1$ to $N$}
\\\quad\quad\quad$\{v^j_l\}_{j=1}^J=f_{v_l}(\{\Sigma_j\}_{j=1}^J,\{R^j_l\}_{j=1}^J)$ // Variance
\\\quad\quad\quad$\{a^j_l\}_{j=1}^J=f_{a_l}(\{\Sigma_j\}_{j=1}^J,\{R^j_l\}_{j=1}^J)$ // Estimate
\EndFor
\\\quad\ \ $t=t+1$ // Increment iteration index.
\\\quad\ \ $\delta=\frac{1}{NJ}\sum_{l=1}^N\sum_{j=1}^J(\widehat{a}^j_l-a^j_l)^2$ // Change in estimate
\EndWhile
\\{\bf Outputs:} Estimate $\a^j,\forall j$
\end{algorithmic}
\end{algorithm}

\begin{figure}[t] 
\centering
\includegraphics[width=8cm]{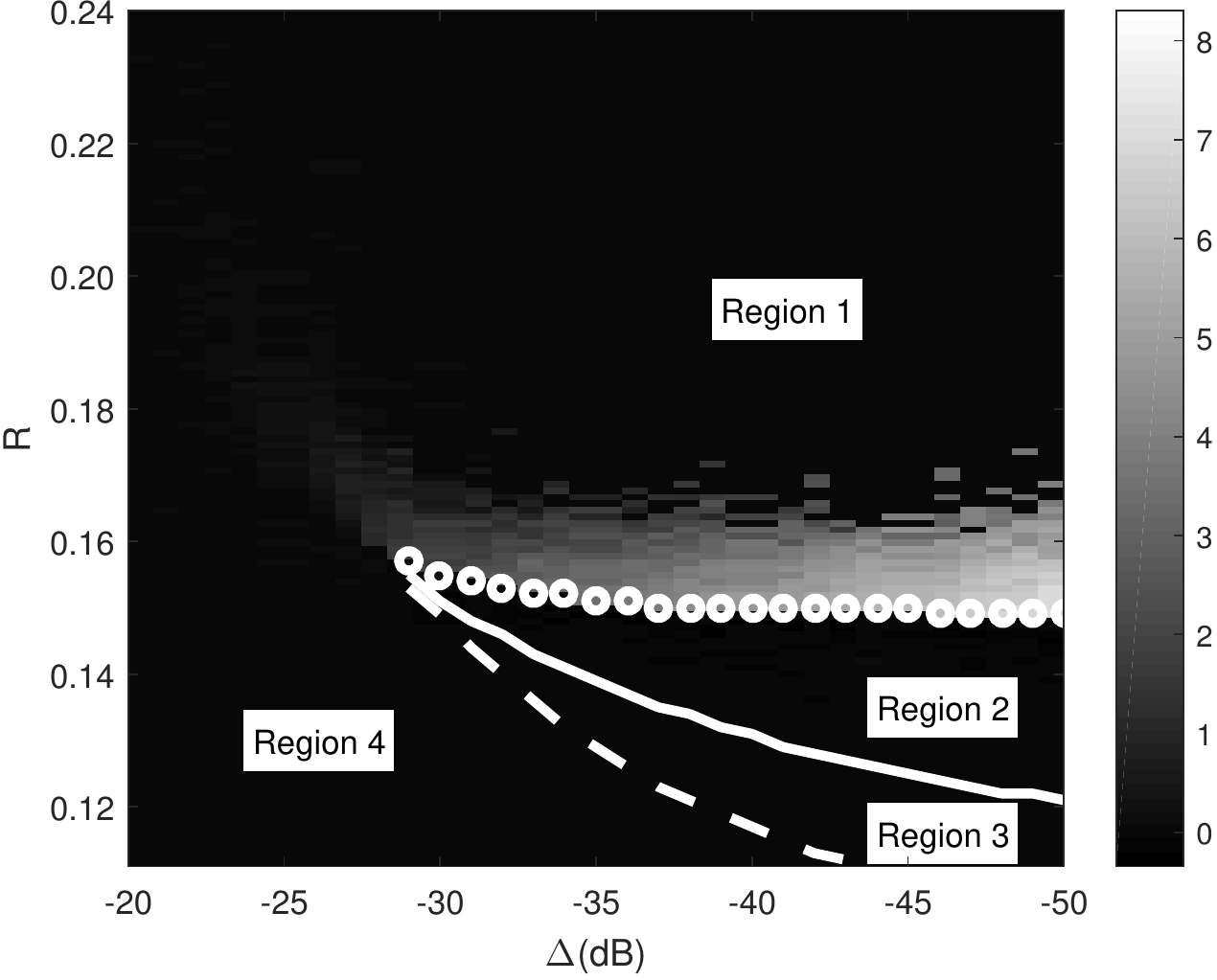}
\caption{AMP simulation results ($\text{MSE}_{\text{AMP}}$) compared to the predicted BP MSE ($\text{MSE}_{\text{BP}}$) with $J=3$ jointly sparse signal vectors. The dashed curve, solid curve, and the curve comprised of little circles correspond to thresholds $R_c(\Delta),\ R_l(\Delta)$, and $R_{BP}(\Delta)$, respectively. Regions 1-4 are also marked. The shade denotes $\ln \l(\frac{\text{MSE}_{\text{AMP}}}{\text{MSE}_{\text{BP}}}\r)$, which we expect to be 0 (completely dark shades) in the entire $R$ versus $\Delta$ plane. The narrow bright band above the BP threshold indicates the mismatch of AMP simulated MSE to the BP predicted MSE.}\label{fig:AMPoverMSE}
\end{figure}

We simulated the signals in~\eqref{eq:jsm} with $J=3$ signal vectors and sparsity rate $\rho=0.1$ measured by a channel~\eqref{eq:MMVmodel} with measurement rate $R\in[0.11,0.24]$ and noise variance $\Delta\in[-20,-50]$ dB. For each setting, we generated 50 signals of length $N=5000$, and the resulting MSE compared to the predicted BP MSE is shown in Fig.~\ref{fig:AMPoverMSE}.\footnote{We simulated both $J$ different measurement matrices $\underline{F}^j$ and $J$ identical $\underline{F}^j$. Both results match the predicted BP MSE, which support our conclusion that the MMSE's of both settings are the same. Fig.~\ref{fig:AMPoverMSE} is with $J$ different $\underline{F}^j$.}
The labels of the thresholds are omitted for brevity. We can see that AMP simulation results match with the predicted MSE of BP and BP phase transition from the replica analysis of Section~\ref{sec:phaseTrans}. Note that there is a narrow band of light shades above the BP threshold, $R_{BP}(\Delta)$ (the top threshold), meaning that the simulated MSE is greater than the predicted MSE; this is due to randomness in our generated signals and channels. Note that we also compared the AMP simulation results to that of the M-SBL algorithm~\cite{YeKimBresler2015}, a widely used algorithm to solve the MMV problem. The M-SBL results were not as good. Indeed, because AMP is often an achievable to the MMSE, other algorithms are expected to provide greater MSE.

\section{Conclusion}\label{sec:conclusion}
We analyzed the minimum mean squared error (MMSE) for two settings of multi-measurement vector (MMV) problems, where the entries in the signal vectors are independent and identically distributed (i.i.d.), and share the same support.
One MMV setting has i.i.d. Gaussian measurement matrices, while the other MMV setting has identical i.i.d. Gaussian measurement matrices. Replica analyses yield identical free energy expressions for these two settings in the large system limit when the signal length goes to infinity and the number of measurements scales with the signal length. Because of the identical free energy expressions, the MMSE's for both MMV settings are identical. By numerically evaluating the free energy expression, we identified different performance regions for MMV where the MMSE as a function of the channel noise variance and the measurement rate behaves differently. We also identified a phase transition for belief propagation algorithms (BP) that separates regions where BP achieves the MMSE asymptotically and where it is suboptimal. Simulation results of an approximated version of BP matched with the MSE predicted by replica analysis. As a special case of MMV, we extended our replica analysis to single measurement vector (SMV) complex CS, so that we can calculate the MMSE for SMV complex CS with real or complex measurement matrices.

\section*{Acknowledgments}
The work in this paper is based in part on preliminary work with Jong Min Kin, Woohyuk Chang, Bangchul Jung, and Jong Chul Ye~\cite{KimChangJungBaronYe2011}.
The authors thank Lenka Zdeborov{\'a} for useful discussions about replica analysis, and Yanting Ma and Ryan Pilgrim for helpful comments. Junan Zhu also thanks Shikai Luo for helpful discussions.

\appendix
This appendix follows the derivation of Barbier and Krzakala~\cite{Barbier2015}, except for some nuances.
Our compressed derivation makes the presentation self-contained.

Plugging~\eqref{eq:XmuNew} and the following identity~\cite{Barbier2015,Krzakala2012probabilistic},
\begin{equation*}
\begin{split}
  &1=\int \exp\Bigg\{-\sij{a=1}{n} \l[\widehat{m}_a\l(m_a NJ-\sij{l=1}{N}(\x_l^a)^T\s_l\r)\r]+\sij{a=1}{n}\Bigg[\\
  &\widehat{Q}_a\!\l(Q_a\frac{NJ}{2}\!-\!\frac{1}{2}\sij{l=1}{N}(\x_l^a)^T\x_l^a\r)\Bigg]\!-\!\sij{1\leq a< b\leq n}{}\Bigg[\widehat{q}_{ab}\Bigg(q_{ab} NJ-\\
  &\sij{l=1}{N}(\x_l^a)^T\x_l^b\Bigg)\Bigg]\!\Bigg\}\!\pij{a=1}{n}dQ_a\  d\widehat{Q}_{a}\  dm_a \ d\widehat{m}_{a}\! \pij{1\leq a<b\leq n}{}\! dq_{ab}\ d\widehat{q}_{ab},
\end{split}
\end{equation*}
into \eqref{eq:EZn1}, we obtain
\begin{equation}\label{eq:EZn3}
\begin{split}
  \mathbb{E}_{\F,\s,\z}&[Z^n]\!=\!(2\pi\Delta)^{-\frac{nMJ}{2}}\bigintsss\! \exp\Bigg[NJ\Bigg(\frac{1}{2}\sij{a=1}{n}\widehat{Q}_aQ_a\\
 & -\frac{1}{2}\sij{\substack{1\leq a,b\leq n\\a\neq b}}{}\widehat{q}_{ab}q_{ab}-\sij{a=1}{n}\widehat{m}_am_a\Bigg)\Bigg]\l[\pij{\mu=1}{M} \mathbb{X}_{\mu}\!\r]\times\\
  &\Gamma^N\pij{a=1}{n}dQ_a\  d\widehat{Q}_{a}\  dm_a\  d\widehat{m}_{a}\!\pij{\substack{1\leq a,b\leq n\\a\neq b}}{}\! dq_{ab}\ d\widehat{q}_{ab},
\end{split}
\end{equation}
where
\begin{equation}\label{eq:Gamma_original}
\begin{split}
&\Gamma=\!\!\!\int\! P(\s_1)\! \l[\pij{a=1}{n}P(\x^a_1)\r]\!\exp\!\Bigg[\!-\frac{1}{2}\sij{a=1}{n}\widehat{Q}_a(\x^a_1)^T\x^a_1+\\
&\frac{1}{2}\sij{\substack{1\leq a,b\leq n\\a\neq b}}{}\widehat{q}_{ab}(\x^a_1)^T\x^b_1+\sij{a=1}{n}\widehat{m}_a(\x^a_1)^T\s_1\Bigg]d\s_1 \pij{a=1}{n}d\x^a_1.
\end{split}
\end{equation}

\begin{figure*}
\begin{equation}\label{eq:PhiJ}
\begin{split}
  \widetilde{\Phi}_J&(m,\widehat{m},q,\widehat{q},Q,\widehat{Q})=\frac{J}{2}(Q\widehat{Q}+q\widehat{q}-2m\widehat{m})-\frac{MJ}{2N}\l[\frac{\rho-2m+\Delta+q}{Q-q+\Delta}+\log(Q-q+\Delta)-\log(\Delta)\r]+\\
  &\int P(\s_1) \l\{ \int \log \l\{ \int P(\x_1)\exp\l[-\frac{1}{2}(\widehat{Q}+\widehat{q})\x_1^T\x_1+\widehat{m}\x^T_1\s_1+
  \sqrt{\widehat{q}}\h^T\x_1\r]d\x_1\r\} \mathcal{D}\h \r\} d\s_1-\frac{MJ}{2N}\log(2\pi\Delta).
\end{split}
\end{equation}
\end{figure*}

\textbf{Further simplification of~\eqref{eq:EZn1}}: The Stratanovitch transform~\cite{Stratanovitch-Wiki} in $J$ dimensions is given by
\begin{equation}\label{eq:strat}
\begin{split}
  &\exp\!\l[\frac{\widehat{q}}{2}\!\sij{\substack{1\!\leq\! a,b\!\leq n\\a\neq b}}{}\!(\x^a_1)^T\!\x^b_1\!\r]\!=\!\pij{j=1}{J}\!\exp\!\l[\frac{\widehat{q}}{2}\sij{\substack{1\leq a,b\leq n\\a\neq b}}{}x^a_{1,j} x^b_{1,j}\r]\\
  &=\pij{j=1}{J}\int\exp\l[\sqrt{\widehat{q}}h_j\sij{a=1}{n}x^a_{1,j}-\frac{\widehat{q}}{2}\sij{a=1}{n}(x^a_{1,j})^2\r]\mathcal{D}h_j\\
  &=\int\exp\l[\sqrt{\widehat{q}}\h^T\sij{a=1}{n}\x^a_1-\frac{\widehat{q}}{2}\sij{a=1}{n}(\x^a_1)^T\x^a_1\r]\mathcal{D}\h,
\end{split}
\end{equation}
where $\h=[h_1,...,h_J]^T$, and the differential $\mathcal{D}h_j=\frac{1}{\sqrt{2\pi}}\operatorname{e}^{-h_j^2/2}d h_j$. With the Stratanovitch transform~\eqref{eq:strat},
we simplify $\Gamma$~\eqref{eq:Gamma_original} as follows,
\begin{equation}\label{eq:Gamma}
\Gamma=\int P(\s_1)\int\l[f(\h)\r]^n \mathcal{D}\h\ d\s_1,
\end{equation}
where
$f(\h)=\int P(\x_1)\operatorname{e}^{-\frac{\widehat{Q}+\widehat{q}}{2}\x_1^T\x_1+\widehat{m}\x_1^T\s_1+\sqrt{\widehat{q}}\h^T\x_1}d\x_1$,
and we drop the super-script $a$ of $\x_1^a$ owing to the replica symmetry assumption~\cite{Krzakala2012probabilistic,krzakala2012statistical}.
In the limit of $n\rightarrow 0$, using another Taylor series $[f(\h)]^n\approx 1+n\log [f(\h)]$, we have $\int [f(\h)]^n \mathcal{D}\h\approx 1+n\int \log [f(\h)] \mathcal{D}\h\approx\operatorname{e}^{n\int \log [f(\h)]\mathcal{D}\h}$, so that $\mathbb{E}\{\int[f(\h)]^n\mathcal{D}\h\}\approx\mathbb{E}\{1+n\int \log [f(\h)] \mathcal{D}\h\}\approx\operatorname{e}^{\mathbb{E}\{n\int \log [f(\h)]\mathcal{D}\h\}}$. Hence, we can approximate~\eqref{eq:Gamma} as
\begin{equation}\label{eq:Gamma1}
\Gamma=\exp\l\{n\int P(\s_1) \int\log [f(\h)] \mathcal{D}\h \ d\s_1\r\}.
\end{equation}
Considering~\eqref{eq:Gamma1}, we rewrite \eqref{eq:EZn3} as
\begin{equation}\label{eq:EZn4}
    \mathbb{E}_{\F,\s,\z}[Z^n]=\int \operatorname{e}^{nN\widetilde{\Phi}_J(m,\widehat{m},q,\widehat{q},Q,\widehat{Q})}
      dm\ d\widehat{m}\ dq\ d\widehat{q}\ dQ\ d\widehat{Q},
\end{equation}
where $\widetilde{\Phi}_J(m,\widehat{m},q,\widehat{q},Q,\widehat{Q})$ is given in~\eqref{eq:PhiJ}.

\textbf{Free energy expression}:
We now substitute~\eqref{eq:EZn4} into~\eqref{eq:replicaTrick}. Assuming that the limits in~\eqref{eq:replicaTrick} commute and that we only evaluate~\eqref{eq:replicaTrick} at optimum points of $\widetilde{\Phi}_J$~\eqref{eq:PhiJ}~\cite{Barbier2015,Krzakala2012probabilistic,krzakala2012statistical}, we have
$\mathcal{F}=\widetilde{\Phi}_J(m^*,\widehat{m}^*,q^*,\widehat{q}^*,Q^*,\widehat{Q}^*)$,
where the asterisks denote stationary points. Next, we calculate the stationary points:
\[\frac{\partial \widetilde{\Phi}_J }{\partial m}=0 \Rightarrow \widehat{m}^*=\frac{R}{Q^*-q^*+\Delta},\]
\[\frac{\partial \widetilde{\Phi}_J }{\partial q}=0 \Rightarrow \widehat{q}^*=R\frac{\Delta+\rho-2m^*+q^*}{(Q^*-q^*+\Delta)^2},\]
\[\frac{\partial \widetilde{\Phi}_J }{\partial Q}=0 \Rightarrow \widehat{Q}^*=R\frac{2m^*-\rho-2q^*+Q^*}{(Q^*-q^*+\Delta)^2},\]
where $R$~\eqref{eq:measurementRate} is the measurement rate. Because we are analyzing the MMSE, we must assume that the estimated prior matches the true underlying prior, which is a Bayesian setting. Thus, $q^*=m^*$ and $Q^*=\rho$~\eqref{eq:auxParamsSet1}. Let $E=q^*-2m^*+Q^*=Q^*-q^*$, then we obtain $\widehat{q}^*=\widehat{m}^*=\frac{R}{E+\Delta}$ and $\widehat{Q}^*=0$. Therefore, we solve for the free energy as a function of $E$ in~\eqref{eq:free_energy3}. Using a change of variables, we obtain~\eqref{eq:free_energy4}, which is a function of $E$. Using~\eqref{eq:auxParamsSet1}, the MSE is
\begin{equation}\label{eq:DandE}
D=E+Q-q=E+\rho/N\overset{N\rightarrow \infty}{\longrightarrow} E.
\end{equation}
Hence, in the large system limit, we can regard the free energy~\eqref{eq:free_energy4} as a function of the MSE, $D$.

\end{document}